\documentclass[conference]{IEEEtran}
\IEEEoverridecommandlockouts
\usepackage{cite}
\usepackage{amsmath,amssymb,amsfonts}
\usepackage{algorithmic}
\usepackage{graphicx}
\usepackage{textcomp}
\usepackage{xcolor}
\def\BibTeX{{\rm B\kern-.05em{\sc i\kern-.025em b}\kern-.08em
    T\kern-.1667em\lower.7ex\hbox{E}\kern-.125emX}}
\begin{document}

\title{Analysis of Security in OS-Level Virtualization
}

\author{
    \IEEEauthorblockN{Krishna Sai Ketha\IEEEauthorrefmark{1}, Guanqun Song\IEEEauthorrefmark{1}, Ting Zhu\IEEEauthorrefmark{1}}
    \IEEEauthorblockA{\IEEEauthorrefmark{1}\textit{Department of Computer Science and Engineering}, \textit{The Ohio State University}, Columbus, USA \\
    Email: ketha.2@buckeyemail.osu.edu, song.2107@osu.edu, zhu.3445@osu.edu}
}

\maketitle

\begin{abstract}
Virtualization is a technique that allows multiple instances typically running different guest operating systems on top of single physical hardware. A hypervisor, a  layer of software running on top of the host operating system, typically runs and manages these different guest operating systems. Rather than to run different services on different servers for reliability and security reasons, companies started to employ virtualization over their servers to run these services within a single server. This approach proves beneficial to the companies as it provides much better reliability, stronger isolation, improved security and resource utilization compared to running services on multiple servers. 

Although hypervisor based virtualization offers better resource utilization and stronger isolation, it also suffers from high overhead as the host operating system has to maintain different guest operating systems. 

To tackle this issue, another form of virtualization known as Operating System-level virtualization has emerged. This virtualization provides light-weight, minimal and efficient virtualization, as the different instances are run on top of the same host operating system, sharing the resources of the host operating system. But due to instances sharing the same host operating system affects the isolation of the instances.

In this paper, we will first establish the basic concepts of virtualization and point out the differences between the hyper-visor based virtualization and operating system-level virtualization. Next, we will discuss the container creation life-cycle which helps in forming a container threat model for the container systems, which allows to map different potential attack vectors within these systems. Finally, we will discuss a case study, which further looks at isolation provided by the containers. 

\end{abstract}

\section{\textbf{Introduction}}
Virtualization is the technique of emulating multiple virtual instances of machines within a single physical hardware, where each virtual machine potentially runs a different operating system. 

Although the concept of virtualization has existed since the 1960s, it has become mainstream since the 1990s \cite{b1} due to the increasing computing demands of companies and due to an increase in the companies opting to run their services in cloud environments, which extensively utilizes virtualization technology.

Companies with the dedicated data centers initially started using it due to the advantages that virtualization provides, such as reliability, fault tolerance, isolation, security, etc. Services provided by these companies are generally hosted on commercial servers within these data centers. As single servers host multiple services, it affects the reliability of the services. For any reason the server running these services crashes, it would have catastrophic effects on the business of the companies. This approach is also not efficient from a security perspective. An attacker who gained access to the host through one of the services, can easily access other services running on the host and disrupt the entire system. 

For improving reliability and security, these services can be separated and each service can be deployed on different servers, with each server running a single dedicated service. This approach improves the reliability and security. But maintaining a large number of servers will be costly and harder to manage. This approach is also computationally not efficient, as some of these services may only need a small portion of computational resources where the remaining resources get wasted. 

Virtualization offers a better approach to run and manage services by providing better reliability, resource utilization, isolation and security. In virtualization, generally, hypervisor (Figure 1), a layer of software running on top of the host operating system, runs and manages guest operating systems. A hypervisor acts as an intermediary between host operating system and guest operating system.
 
\begin{figure}
    \centering
    \includegraphics[width=8cm]{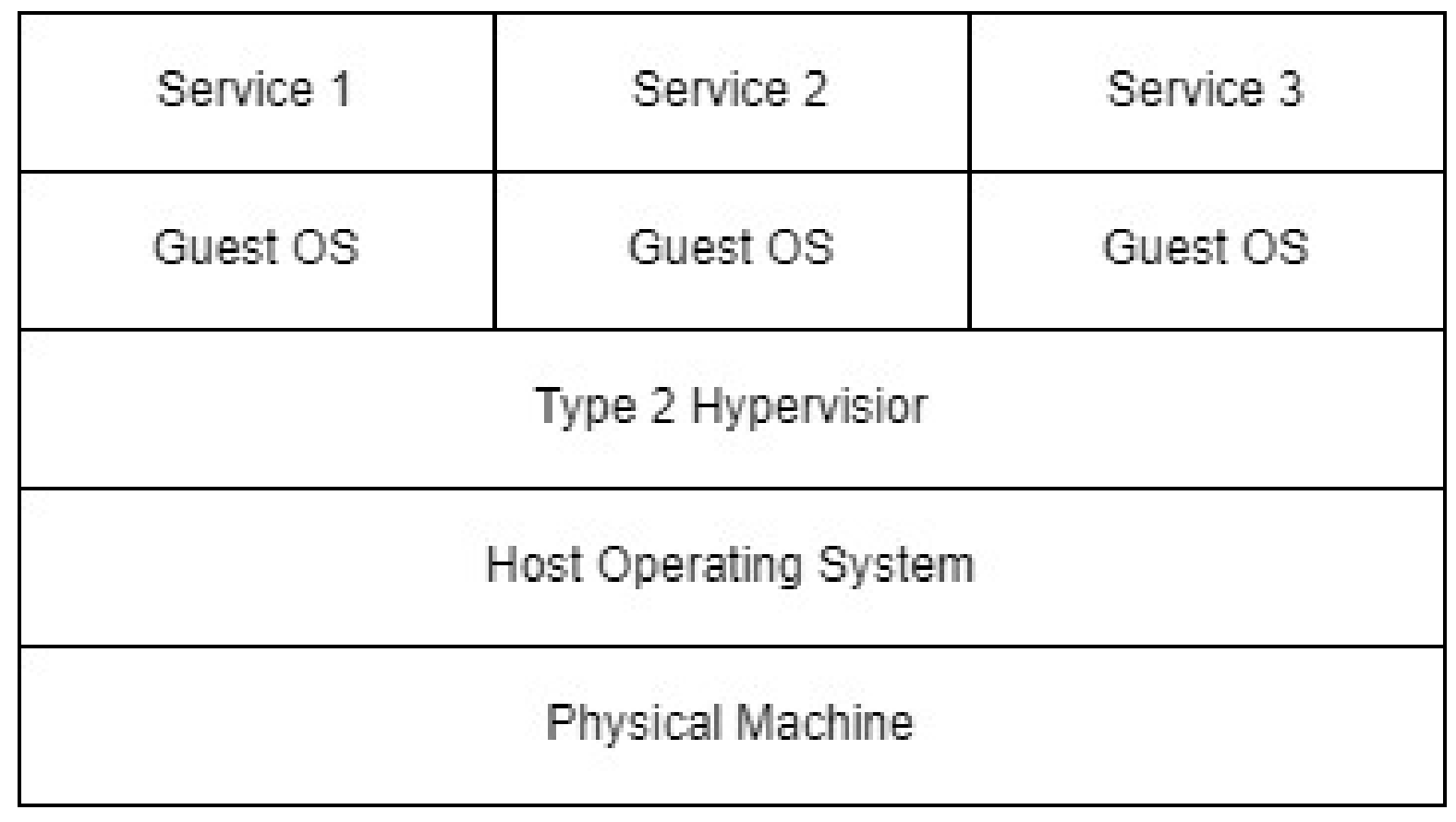}
    \caption{Type 2 - Virtualization}
    \label{fig:virtualization}
\end{figure}

Running services with virtual machines provides reliability as any crashes only disrupts the service within the virtual machine, where the rest of the system works uninterrupted. It also improves security, as any intruder gaining access to the system, will be contained to the respective virtual machine, providing stronger isolation. It also provides better resource utilization, as multiple services can be run on a single physical machine. 

Apart from reliability, isolation, security and resource utilization, virtualization also improves developer experience. Consider a scenario where two services require two different dependencies, each of these dependencies compatible with two different types of operating systems. In such a scenario it would be difficult to run these services on a single machine and it requires two different machines. With virtualization, we can have two different operating systems, configured as per the needs of the two services, running the services simultaneously on top of a single physical machine. It is also necessary for developers to test the services with different operating systems to make sure the services are working without any issues on different ranges of machines. 

Virtualization also supports check-pointing and migration. Check-pointing is the process of periodically storing the state of a system. With check-pointing, it will be possible to restore to the last state of the system. This would decrease the effects of the crash on the systems. Migration is the process of migrating the system from one system to another system. Consider a scenario where a developer team is working on a service. But as the service grows it would require more resources to run. In this scenario it would be beneficial to migrate the service from one system to another system to compensate for the resources. 

While virtualization offers numerous benefits, it also presents certain drawbacks. Primary drawback of virtualization is additional resource overhead. Each virtual machine runs a dedicated guest operating system within, which increases the load on the host system, which has to manage the service and an entire operating system.

To address this challenge, another virtualization approach known as Operating system-level virtualization is utilized. The virtualization we discussed up until now is also known as hypervisor-based virtualization. In Operating system-level virtualization (Figure 2), different virtual instances run on top of the host operating system unlike hypervisor-based virtualization, where multiple instances run within different guest operating systems. 

\begin{figure}
    \centering
    \includegraphics[width=8cm]{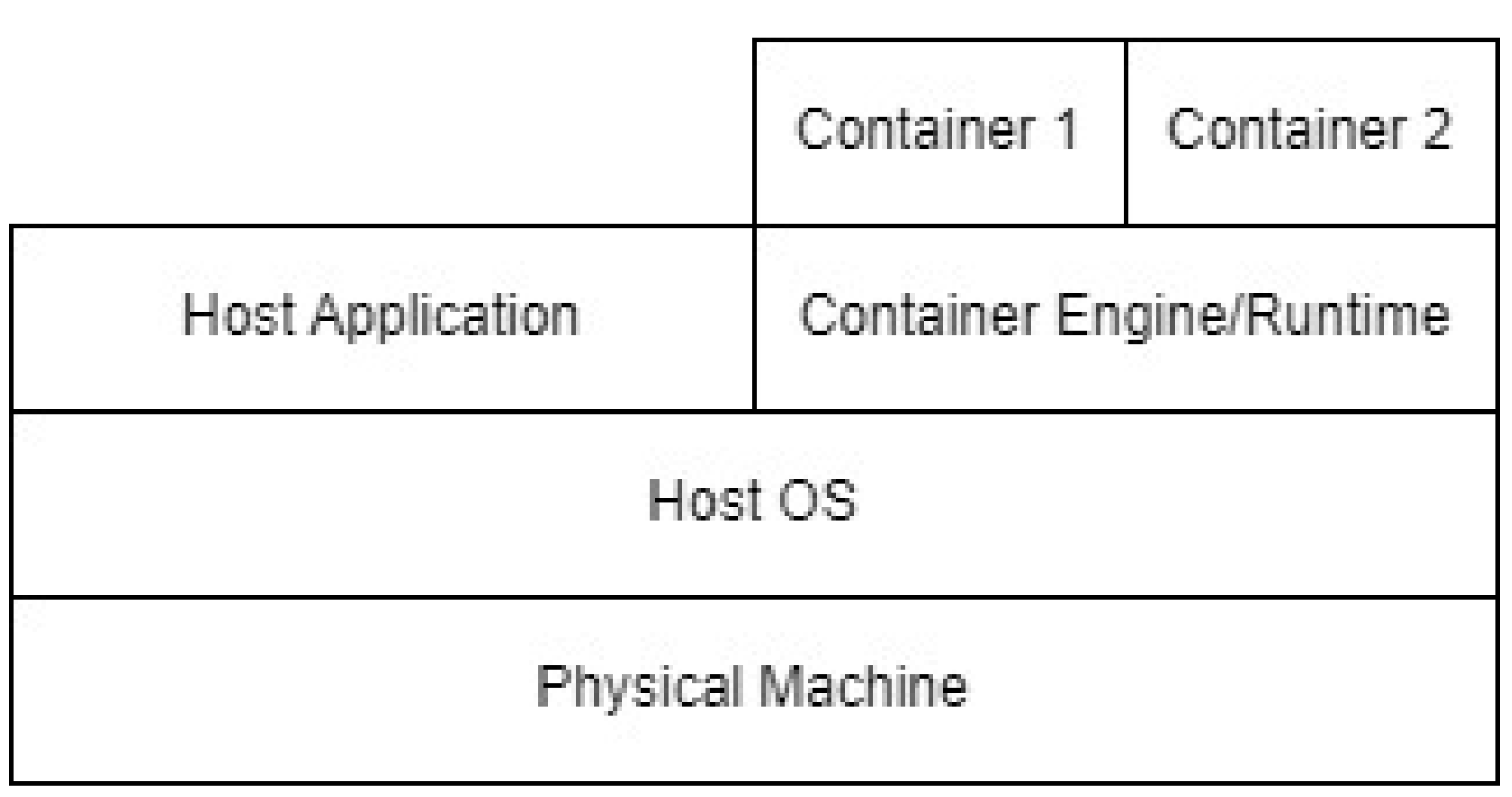}
    \caption{OS-Level Virtualization}
    \label{fig:os-level virtualization}
\end{figure}

While operating system-level virtualization instances may seem like different isolated systems, they all share the same underlying guest operating system and also the resources of the guest operating system. Due to this operating system-level virtualization offers a much more light-weight, minimal and efficient virtualization compared to hypervisor-based virtualization. 

Although operating system-level virtualization offers a minimal overhead and efficient virtualization solution, it might not offer better isolation compared to hypervisor-based virtualization, due to the fact that virtual instances share the same operating system and its resources. \cite{b2}

In this study, we aim to investigate the isolation and security provided by operating system-level virtualization. Specifically, we will examine the process through which operating system-level virtualization creates and manages virtual instances. Additionally, we will develop a container threat model to identify potential attack vectors within containerized systems, and explore potential solutions to safeguard containers from such threats.

The rest of the paper is structured as follows: Section 2 discusses container creation life-cycle. Section 3 talks about a container threat model to identify potential attack vectors in a containerized system. Section 4 provides an analysis of two case studies examining actual vulnerabilities discovered in Docker. Section 5 presents a brief conclusion.

\section{\textbf{Container Creation Life-cycle}}
As mentioned earlier, OS-Level virtualization implements multiple user-level instances on top of a single host operating system. There are various popular implementations of OS-level virtualization, such as FreeBSD Jails, Solaris Zones, and Docker Containers. For this paper, we will be looking at the Docker Containers to understand the isolation and security in the OS-level virtualization. 

To understand the isolation and security offered by operating system-level virtualization, we began by analyzing how containers are built within Docker, transferred to and from public registries, and run within Docker, as these steps constitute the majority of the container creation life-cycle. Later, Based on the container creation model, we prepared a threat model of the docker to further analyze the security and isolation of the system.

Threat modeling is the systematic identification of potential threats and vulnerabilities existing in a system. To create a threat model of a system, we examine the components of the system and consider various modes of attack existing within a system. \cite{b3} A threat model can help in understanding areas of vulnerabilities existing within the system. 

Docker provides most of the tools required to build, manage, and run containers. Additionally, it supports the integration of third-party tools and services to manage containers. Major tools that are part of the container creation life-cycle include Docker Engine, Docker Daemon, Containerd, and runc \cite{b4}. Before understanding how containers are created, first, let’s look at each of the components that ships with docker.

Docker Engine: As per the Docker official website, “Docker Engine is an open source containerization technology for building and containerizing your applications.”. Primarily, Docker Engine facilitates the creation and management of containers. Docker Engine essentially contains Docker Command Line Tool Interface intended for the user to interact with docker to create and manage containers, API's for programs running inside docker containers to interact with docker daemon or dockerd. 

Docker Daemon: Dockerd or Docker Daemon is a process that is running in the background listening for the commands from the docker. As we will see later it acts as an interface between docker cli and container runtime.

Containerd: Containerd is a high-level container runtime system. It essentially provides an environment for transferring and managing images, container life-cycle management, storage and networking.

Runc: Runc is a lightweight, secure and low-level container runtime system which interacts with the underlying operating system for creating, running and managing containers.

\begin{figure}
    \centering
    \includegraphics[width=3.5cm]{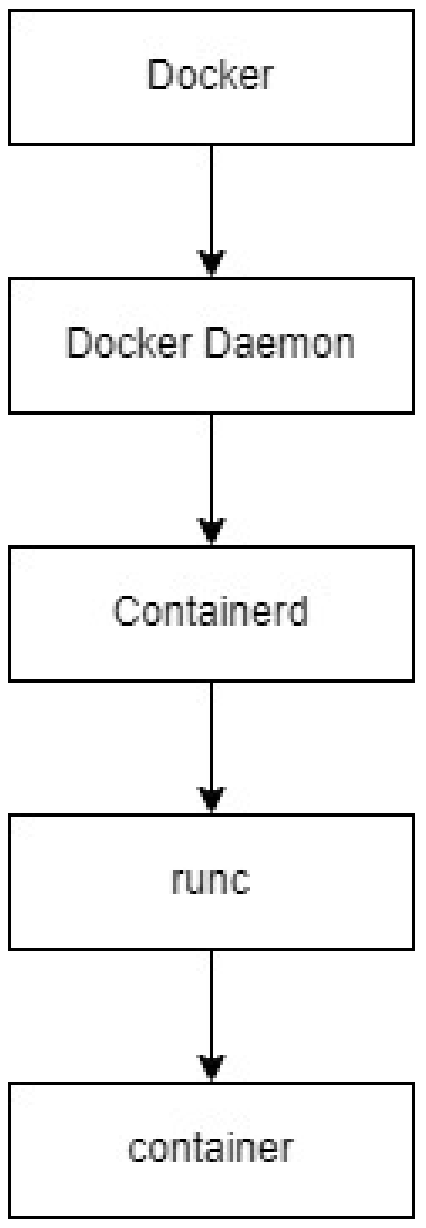}
    \caption{Container Creation Life-cycle}
    \label{fig:container creation lifecycle}
\end{figure}

In order to create a container, first, a dockerfile has to be created with all the necessary configuration to build the docker image. The docker image is then pushed onto a public repository. The docker pushed on to the public repository can be pulled to any remote system and used for building a docker container. 

A system administrator  creates a container by initiating a “docker build” command through the docker command line interface. The Docker CLI takes the request and calls docker daemon. Docker daemon, further process the request to containerd to create a container. Containerd receives the request from the docker daemon, pulls the actual image and passes it to the runc. Runc, which directly interacts with the underlying operating system, creates and manages the container based on the image configuration pulled.
    
    \begin{figure*}
        \centering
        \centerline{\fbox{\includegraphics[width=1\textwidth]{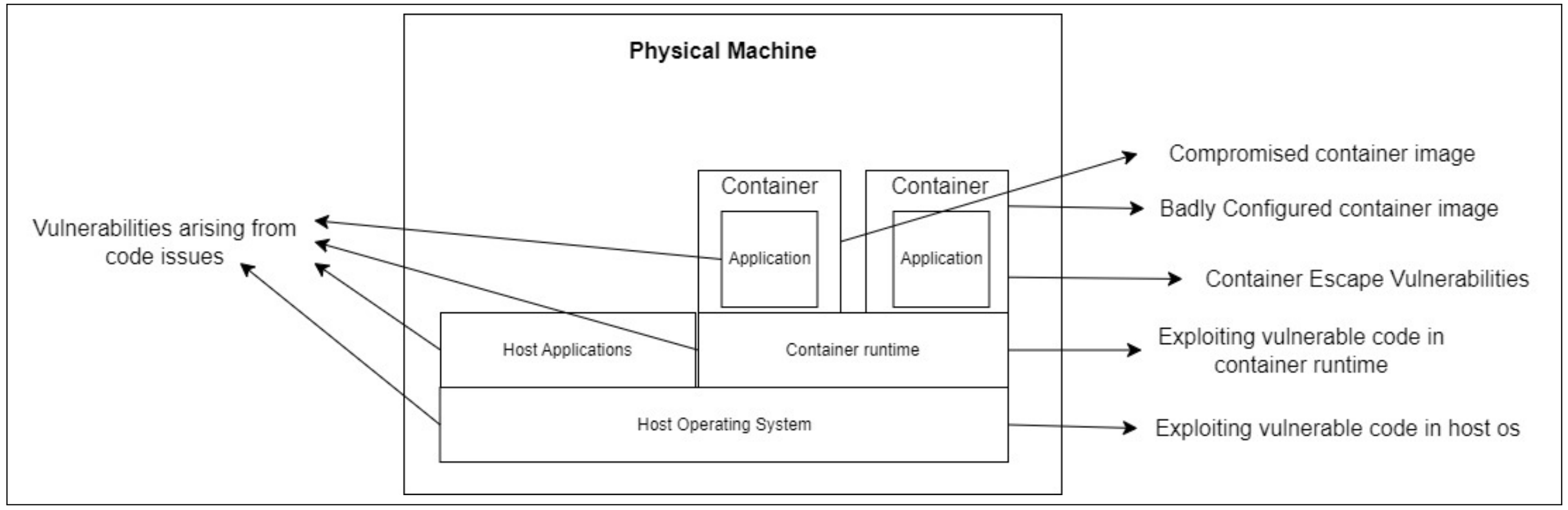}}}
        \caption{Container Threat Model}
        \label{fig}
    \end{figure*}
    
\section{\textbf{Container Threat Modeling}}

By examining the container creation lifecycle, we can identify (Figure 4) the majority of potential attack vectors present at each stage of the container lifecycle.

In a container setup, attackers typically attempt to gain access to a system either externally or internally. External attacks often involve remote access to the container system, allowing attackers to coordinate attacks through a network connected to the container. Alternatively, attackers may already be within the container as users and initiate attacks from within. This may involve attempting to run commands on the running container to escalate privileges or executing malicious code within the container, causing the system to perform unintended actions.

After examining the potential attack vectors within containers, these attacks can primarily be classified into two categories: those originating over the container network and those exploiting local vulnerabilities. Moreover, these potential attack vectors can be further classified based on the origin of vulnerabilities within the container system. These are:

\begin{enumerate}
    \item \textbf{Manipulating container images on repository:} Container images are generally stored in Docker repositories which are later pulled to create and run containers. But, the issue arises with image-integrity. The attacker can find the image and tamper the image which can give the attacker access to the container.
    \item \textbf{Vulnerabilities with application code:} An attacker can exploit the vulnerabilities that exist within the application code and third-party dependencies used by the application code. There are thousands of vulnerabilities existing in the third-party dependencies which are not patched regularly. The attacker can find these vulnerabilities and can compromise the container.
    \item \textbf{Poorly configured container images:} It is possible to poorly configure the container images, for instance providing the container more privileges than it requires.
    \item \textbf{Poorly configured containers:} It is possible to run containers by pulling images from the public directory, which are configured by the attacker consisting of malicious code which gains access to the container. 
    \item \textbf{Host Vulnerabilities:} Hosts operating systems running the containers can have multiple vulnerabilities, which can be leveraged by the attacker to gain access.
    \item \textbf{Information passing within the system:} Containers often share information with other containers that share the same host operating system. An attacker can track this information which can be further used to gain access to the containers.Information passing within the system: Containers often share information with other containers that share the same host operating system. An attacker can track this information which can be further used to gain access to the containers.
    \item \textbf{Container escape vulnerabilities:} These attacks are generally originated due to the vulnerabilities in the container runtime systems like containerd or runc. The attacker uses the vulnerability existing in one of these systems and tries to escape the container to gain access to the underlying host operating system. These attacks are also known as runcescape.
    \item \textbf{Container communicating over insecure networks:} It is possible that the attacker can track the communication between the containers and use the information to launch attacks over the containers.
    \item \textbf{Uncontrolled Resource Consumption:} An attacker could run a container containing malicious scripts which would allow the container to take up more resources than it needs, allowing the other containers running within the host operating system to starve.
\end{enumerate}

For this paper, we will be particularly analyzing the vulnerabilities that originate locally.

\section{\textbf{Case studies}}
\subsection{\textbf{Runc Container Breakout}}
Runc Container Breakout, a security vulnerability originating due to a flaw in the runc container runtime system, allows an attacker to gain access to the underlying host file system, which the attacker can further utilize to gain privileges to the underlying host operating system \cite{b5}. The vulnerability was first reported on January 31st, 2024, documented as CVE-2024-21626. The range of runc versions affected include v1.0.0-rc93 to v1.1.11, containerd v1.4.7 to v1.6.27 and v1.7.0 to v1.7.12 and docker version <v25.0.2.

Generally, a container is another process from the perspective of the host operating system. But to separate the containers from the host operating and other containers. A container is mounted with a separate filesystem as its root filesystem using chroot. “chroot” is a linux utility which modifies the working root directory for the current process. It limits the process access from the rest of the filesystem. 

\begin{figure}
    \centering
    \includegraphics[width=8cm]{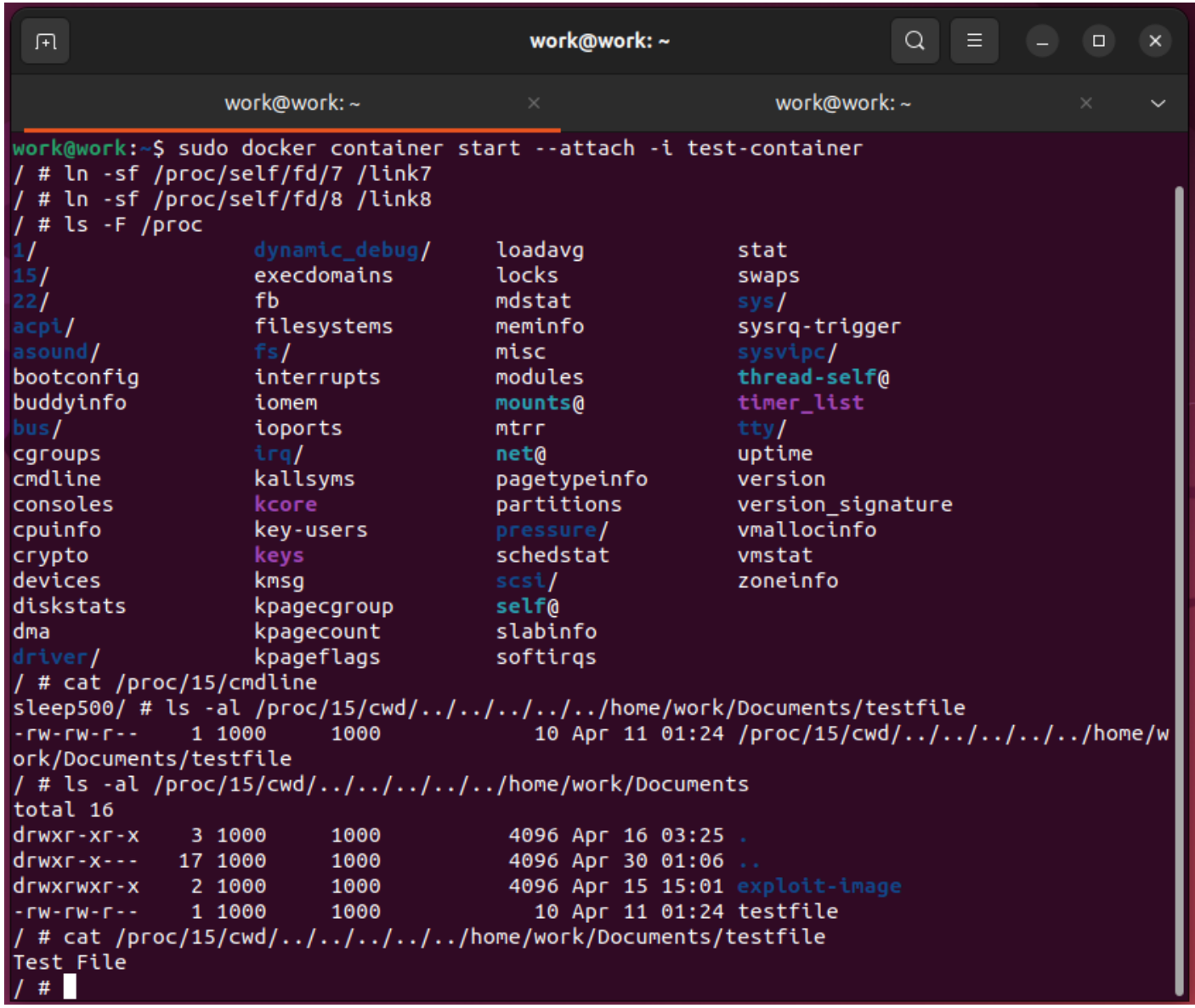}
    \caption{Runc Container Breakout and Reading files within host operating system}
    \label{fig:container breakout}
\end{figure}

\begin{figure}
    \centering
    \includegraphics[width=8cm]{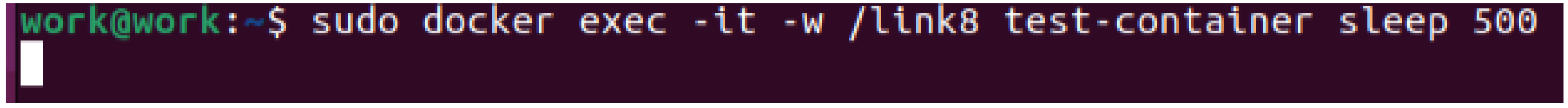}
    \caption{Executing Docker exec command which mounts the host filesystem in current working directory}
    \label{fig:container breakout}
\end{figure}

In the above attack, we created symlinks to file descriptors 7 and 8 (Figure 4). The /proc/self/fd contains the file descriptors of current processes. When the “sudo docker exec -it -w /foo <container-name> sleep 500” (Figure 5) command is executed, runc mounts the host filesystem on to the container which allows the attacker to gain access to the filesystem and allows to read the files within the host filesystem.

The Runc container breakout is classified as a high severity attack due to the fact the attacker doesn’t need to run any malicious code, requiring high privileges to compromise the system.

\subsection{\textbf{Dirty Pipe Vulnerability}}
Dirty Pipe Vulnerability, is a vulnerability that was found in the linux kernel host operating system. This vulnerability allows the attacker to gain access to arbitrary read-only files like /etc/passwd and modify them which provides elevated access to the host operating system \cite{b6}. The vulnerability was first reported on March 7th, 2022, documented as CVE-2022-0847. The range of linux versions affected include >v5.8.0.

 A vulnerability was identified in the Linux kernel, specifically in the "flags" member of the new pipe buffer structure. This flaw occurred due to improper initialization in the “copy-page-to-iter-pipe” and “push-pipe” functions, potentially resulting in the inclusion of outdated values. An unprivileged local user could exploit this vulnerability to modify pages in the page cache associated with read-only files, thereby escalating their privileges on the system.

 \section{Future Work}
 In recent years, technology has gained a lot of progress in various fields especially in the direction of security \cite{wire2, 10125074,285483,10.1145/3395351.3399367}, artificial intelligence \cite{10.1145/3460120.3484766,9709070,9444204,ning2021benchmarkingmachinelearningfast,8832180,8556807,8422243,chandrasekaran2022computervisionbasedparking,iqbal2021machinelearningartificialintelligence,pan2020endogenous}, and the Internet of Things (IoT) \cite{wire1,wire3,10017581, 9523755,9340574,10.1145/3387514.3405861,9141221,9120764,10.1145/3356250.3360046,8737525,8694952,10.1145/3274783.3274846,10.1145/3210240.3210346,8486349,8117550,8057109,https://doi.org/10.1155/2017/5156164,10189210}. The trend of the future is to combine various advanced and excellent technologies in order to find more efficient strategies. In this case, we will focus on integrating AI-driven threat detection and secure communication protocols to enhance the resilience of containerized systems against evolving vulnerabilities. Additionally, optimizing container networking for wireless environments and improving resource allocation through system-level innovations will be crucial to ensure both performance and security in dynamic and distributed infrastructures.

\section{\textbf{Conclusion}}
This section discusses the techniques and methods to harden security within the container environments. Although continuous effort has been put into hardening the security of the containers, attackers always try to find vulnerabilities within the containerized systems and exploit these vulnerabilities. It doesn’t matter how battle-tested these systems are, there is always a chance of an unnoticed vulnerability existing within the systems which the attacker finds and exploits. 

For instance, the vulnerability mentioned earlier in the case study, CVE-2024-21626: runc container breakout, isn’t the first of its kind. Many prior vulnerabilities such CVE-2019-5736 is of the same kind, where the attacker exploits the vulnerability within runc to gain privileges to the host operating system and execute devastating attacks.

In order to keep the container safe from attacks it is alway suggested to follow best security practices. First, always check for the container's images being pulled from the public repositories, and scan these for potential vulnerabilities. Next, make sure you haven’t accidently provided unnecessary privileges to the images. An attacker potentially gaining access to the container can easily use these containers to perform malicious attacks over other containers. Next, try to keep the container engines, run-times, tools and third party tools as updated as possible. As mentioned before, it doesn't matter how battle-tested these systems are, some form of vulnerabilities always exist in these systems. These companies always try to release patches to these systems, so it’s always better to update the tools to the current recommended version. Next, make sure your dependencies don't contain any vulnerabilities. Most of the time even though we implemented all the precautions to secure the containers. They can be compromised due to the vulnerabilities of the third party dependencies. Next, try to pass information to other containers over secure connections, most of the time attackers monitor the communication channels to intercept the communications between the containers and use them to gain access to the systems. Finally, try to scan the containers using security scripts to check for any vulnerabilities.


\bibliographystyle{ieeetr} 
\bibliography{zhu}
\end{document}